\newif\ifdouble

\ifdouble
    \documentclass[journal, 10pt, twoside]{IEEEtran}
\else
    \documentclass[journal, 12pt, draftcls, onecolumn, twoside]{IEEEtran}
\fi

\usepackage[T1]{fontenc} 

\usepackage{cite}

\ifCLASSINFOpdf
\usepackage[pdftex]{graphicx}
\graphicspath{{./diagrams/}{./images/}{./bio/}}
\DeclareGraphicsExtensions{.pdf,.jpeg,.png,.eps}
\usepackage{epstopdf}
\else
\fi
\usepackage{amsmath}
\usepackage{algorithmic}

\usepackage{algorithm}

\makeatletter
\newcommand\fs@norules{\def\@fs@cfont{\bfseries}\let\@fs@capt\floatc@ruled
	\def\@fs@pre{}%
	\def\@fs@post{}%
	\def\@fs@mid{\kern3pt}%
	\let\@fs@iftopcapt\iftrue}
\makeatother
\restylefloat{algorithm}

\usepackage{array}

\usepackage[caption=false,font=footnotesize]{subfig}
\usepackage{url}

\newcommand{\specialcell}[2][c]{%
	\begin{tabular}[#1]{@{}c@{}}#2\end{tabular}}

\usepackage{color}
\usepackage{amsfonts}
\usepackage{amssymb}
\usepackage{tikz}
\usepackage{hyperref}

\usepackage{tikz}
\usetikzlibrary{patterns}
\usetikzlibrary{backgrounds}
\usetikzlibrary{calc}
\usetikzlibrary{shapes}
\usetikzlibrary{fit}
\usetikzlibrary{chains}
\usetikzlibrary{arrows}
\usepackage{multirow}
\usepackage{tkz-graph}
\usetikzlibrary{intersections}
\usepackage{tkz-euclide}
\usetkzobj{all}
\usepackage{multirow}
\usepackage{tkz-graph}

\tikzstyle{block} = [draw, rectangle, minimum height=2em, minimum width=0.5em]
\tikzstyle{sum} = [draw, circle, node distance=1cm, fill=white]
\tikzstyle{input} = [coordinate] \tikzstyle{output} = [coordinate]
\tikzstyle{pinstyle} = [pin edge={to-,thin,black}]

\usepackage[normalem]{ulem}

\begin{document}
	%
	\title{Iterative Decision Feedback Equalization Using Online Prediction}
		%
	%
	%
	
	\author{Serdar~\c{S}ah\.{i}n*,~
		Antonio~Maria~Cipriano,~
		Charly~Poulliat,
		and~Marie-Laure~Boucheret%
		\thanks{Manuscript sent to IEEE. \emph{(Corresponding author: Serdar \c{S}ahin.})}
        \thanks{S. \c{S}ah\.{i}n is with Thales Communications and Security, 92230 Gennevilliers, France, and also with IRIT/INPT-ENSEEIHT, 31000 Toulouse, France (e-mail: serdar.sahin@thalesgroup.com).}%
        \thanks{A. M. Cipriano is with Thales Communications and Security, 92230, Gennevilliers, France (e-mail: antonio.cipriano@thalesgroup.com).}%
        \thanks{C. Poulliat and M.-L. Boucheret are with IRIT/INP Toulouse-ENSEEIHT, 31000, Toulouse, France (e-mails: charly.pouillat@enseeiht.fr, marie-laure.boucheret@enseeiht.fr).}%
        \thanks{Digital Object Identifier XXXXXXXXXXXXX.}
	}

	
	%
	%

	\markboth{Draft for IEEE - July 2019}%
	{\c{S}ah\.{i}n \MakeLowercase{\textit{et al.}} -  Iterative Decision Feedback Equalization Using Online Prediction}
	%



	\maketitle
	
\begin{abstract}
    In this article, a new category of soft-input soft-output (SISO) minimum-mean square error (MMSE)  finite-impulse response (FIR) decision feedback equalizers (DFEs) with iteration-wise static filters (i.e. iteration variant) is investigated.
    It has been recently shown that SISO MMSE DFE with dynamic filters (i.e. time-varying) reaches very attractive operating points for high-data rate applications, when compared to alternative turbo-equalizers of the same category, thanks to sequential estimation of data symbols \cite{sahinCipriano_2018_DFEEP}.
    However the dependence of filters on the feedback incurs high amount of latency and computational costs, hence SISO MMSE DFEs with static filters provide an attractive alternative for computational complexity-performance trade-off. 
    However, the latter category of receivers faces a fundamental design issue on the estimation of the decision feedback reliability for filter computation.
    To address this issue, a novel approach to decision feedback reliability estimation through online prediction is proposed and applied for SISO FIR DFE with either a posteriori probability (APP) or expectation propagation (EP) based soft feedback.
    This novel method for filter computation is shown to improve detection performance compared to previously known alternative methods, and finite-length and asymptotic analysis show that DFE with static filters still remains well-suited for high-spectral efficiency applications.
\end{abstract}


	%
	\IEEEpeerreviewmaketitle

\section{Introduction}
\label{sec:introduction}
{J}{oint} detection and decoding through iterative exchange of extrinsic information between a soft-input soft-output (SISO) detector and a SISO decoder can achieve near capacity performance with a well-designed coding scheme.
In particular, turbo equalization seeks to provide robust high-throughput links over strongly frequency-selective channels, whose frequency responses' incorporate spectral nulls.

However, unlike SISO finite-impulse response (FIR) minimum mean squared error (MMSE) linear equalizers (LE) \cite{tuchler_turbo_2011}, practical SISO MMSE FIR DFE structures have not been thoroughly investigated, and only gathered attention in recent years \cite{lou_soft_2014,tao_2016_low,sahinCipriano_2018_DFEEP}. 
In this article, a novel filter computation approach for such structures is proposed, through a predictive estimation of the decision feedback reliability.
In the following, we only consider SISO MMSE FIR receiver structures.

A widespread and widely accepted nomenclature for categorizing such equalizers has not been established in the literature.
In our view, the work in \cite{jeong_2011_low_complexity_bidir} provides an accurate categorization based on the occurrence of SISO adaptive filter updates with prior information.
This nomenclature is attractive as it is directly related to the assumptions used for the derivation of the equalizer, and it also gives some insights on both the decoding performance and the computational complexity. 
Time varying (TV) FIR filters are updated at each single symbol, by fully exploiting prior information, and they are well-suited for doubly-selective channels.
Iteration varying (IV) FIR filters are static and updated only at the beginning of each turbo iteration, by using the knowledge of overall quality of the feedback, thus reducing the involved computational costs. 

The nature of feedback ``decisions'' (or rather ``estimates'') also impacts the DFE error-rate performance. 
Hard decisions can be taken, as in conventional DFE \cite{belfiore_79_DFE}, but this results in a significant amount of error propagation and unpredictable behavior \cite{tuchler_turbo_2002}, unless complex heuristics are used \cite{jeong_turbo_2010}.
Alternatively, a widespread category of DFE with soft feedback use a posteriori probability (APP) distributions, due to its relative simplicity and fair performance \cite{lopes_soft-feedback_2006,lou_soft_2014, tao_2016_low}. 
Finally, soft feedback based on expectation propagation (EP) \cite{minka_2001_expectationpropagation} improves the performance of TV DFE \cite{sahinCipriano_2018_DFEEP}, in addition to being more predictable, due to relatively lowered correlations with the equalized estimates \cite{liWuTao_2019_perfAnalysisAndImprovemetForVAMPFDE}.

The design of optimum IV DFE receivers is however non-trivial. Indeed, static filters should depend on the decision feedback reliability, and the decision feedback naturally depends on the filters.
As a consequence, there does not exist closed-form expression of the optimal filter due to this non-linear ``chicken-and-egg'' inter-dependence.

IV DFE proposals in the literature use a variety of sub-optimal heuristics for filter computation \cite{lopes_soft-feedback_2006, lou_soft_2014, tao_2016_low}. 
For DFE with hard decisions, the conventional approach is to assume a perfect feedback, causing error propagation and performance degradation in real operation \cite{tuchler_turbo_2002}. 
Using soft APP decisions while still assuming perfect feedback partially mitigates error propagation, as soft symbols' magnitudes scale down with unreliability \cite{balakrishnan_mitigation_1999}. 
The first reference to incorporate APP soft feedback reliability in filter computations is the receiver proposed in \cite{lopes_soft-feedback_2006} for the special case of binary-phase-shift-keying (BPSK) modulation.
The direct dependencies between soft symbols and log-likehood ratios (LLRs) for the BPSK constellation enable the use of a tractable density evolution on the APP LLR distribution, given a prior LLR distribution from the decoder.
This property is used by the BPSK receiver of \cite{lopes_soft-feedback_2006} to estimate the decision feedback reliability.
However, this scheme cannot be directly generalized to high-order constellations; hence \cite{lou_soft_2014} proposed a receiver that implements a LE at the first turbo iteration, and then it uses previous turbo iteration's demapper APP LLRs to estimate soft-symbol statistics for IV DFE. 
Note that this approach is only possible with Gray mapped constellations. 
More recently, \cite{tao_2016_low} proposed pre-equalization with LE over a few symbols, and then to use APP distribution of these symbols to estimate APP soft-feedback statistics for IV DFE.

Considering the previous developments, our contributions can be summarized as follows:
\begin{itemize}
    \item An IV DFE with soft APP feedback based on online prediction is formulated. 
    The technique in \cite{lopes_soft-feedback_2006} is extended with models from the field of semi-analytic performance prediction \cite{visoz_semi-analytical_2010, sabbaghian_2009_analyticalapproach, yuan_2008_evolutionanalysis}, used both for physical layer abstraction or link layer adaptation.
    \item A low complexity IV DFE with soft EP-based feedback is proposed, as an IV extension of \cite{sahinCipriano_2018_DFEEP}, by removing the bias from predicted APP estimates' reliability.
    \item A novel approach for estimating soft decision feedback reliability is proposed, based on online binary and symbol-wise semi-analytic performance prediction.
    \item The accuracy and the convergence of these prediction schemes are evaluated and the performance of IV equalizers are compared. To our knowledge, the accuracy of different IV DFE heuristics is not compared elsewhere, despite their direct impact on DFE error propagation.
\end{itemize}

The remainder of this paper is organized as follows. Section~\ref{sec:modelling} presents the system model and the general structure of proposed IV DFE receivers. The involved semi-analytic prediction scheme is developed and analyzed in Section~\ref{sec:abstraction}. 
Finally, before concluding, the proposed equalizers with online prediction are numerically analyzed in Section~\ref{sec:perfos}, in the finite-length and the asymptotic regimes.

\begin{figure}[!t]
	\centering
    
	\includegraphics[width=0.48\textwidth]{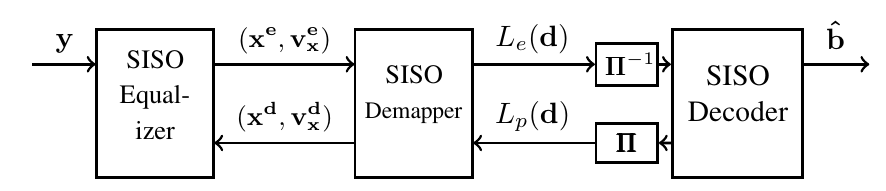}
	\caption{Iterative Detection and Decoding of BICM Signals.}
	\label{fig_0_iter_recv_model}
\end{figure}

\subsection*{Notations}

Bold lowercase letters are used for vectors: let $\mathbf{u}$ be a $N \times 1$ vector, then $u_n, n=1,\dots,N$ are its entries, unless specified otherwise. 
Capital bold letters denote matrices: for a $N \times M$ matrix $\mathbf{A}$, $[\mathbf{A}]_{n,:}$ and $[\mathbf{A}]_{:,m}$ respectively denote its $n^\text{th}$ row and $m^\text{th}$ column, and $a_{n,m}=[\mathbf{A}]_{n,m}$ is the entry $(n,m)$. 
$\mathbf{I}_N$ is the $N\times N$ identity matrix, $\mathbf{0}_{N,M}$ and $\mathbf{1}_{N,M}$ are respectively all zeros and all ones $N\times M$ matrices. $\mathbf{e}_n$ is the $N\times 1$ indicator whose only non-zero entry is $e_n=1$.
Operator $\mathbf{Diag}(\mathbf{u})$ denotes the diagonal matrix whose diagonal is defined by $\mathbf{u}$.
$\mathbb{R}, \mathbb{C}$, and $\mathbb{F}_k$ are respectively real, complex and $k^\text{th}$ order Galois fields.

Let $x$ and $y$ be two random variables, $\mu_x=\mathbb{E}[x]$ is the expected value, $\sigma_x^2=\text{Var}[x]$ is the variance and $\sigma_{x,y}=\text{Cov}[x,y]$ is the covariance. The probability of the discrete random variable $x$ taking the value $\alpha$ is $\mathbb{P}[x=\alpha]$.
For random vectors $\mathbf{x}$ and $\mathbf{y}$, we define  $\pmb{\mu}_\mathbf{x}=\mathbb{E}[\mathbf{x}]$ and covariance matrices $\mathbf{\Sigma}_{\mathbf{x},\mathbf{y}}=\mathbf{Cov}[\mathbf{x,y}]$ and  $\mathbf{\Sigma}_{\mathbf{x}}=\mathbf{Cov}[\mathbf{x,x}]$.
$\mathcal{CN}(\mu_x,\sigma_x^2)$ denotes the circularly-symmetric complex Gaussian distribution of mean $\mu_x$ and variance $\sigma_x^2$.

\section{System Model}\label{sec:modelling}
\subsection{Single Carrier BICM Transmission}

Single carrier transmission using a bit-interleaved coded modulation (BICM) scheme is considered. 

Let $\mathbf{b}\in\mathbb{F}_2^{K_b}$ be a $K_b$-bit information packet.
$\mathbf{b}$ is encoded and then interleaved into a codeword $\mathbf{d}\in \mathbb{F}_2^{K_d}$. 
A memoryless modulator $\varphi$ then maps $\mathbf{d}$ to the symbol block $\mathbf{x}\in \mathcal{X}^K$, where $\mathcal{X}$ is the $M^\text{th}$ order complex constellation, with zero mean and average power $\sigma_x^2=1$, and with $Q=\log_2 M$. 
The $Q$-word associated to $x_k$ is denoted $\mathbf{d}_k=[\mathbf{d}]_{Q(k-1)+1:Qk}$, and both $d_{k,q}$ and $\varphi_q^{-1}(x_{k})$ denote the value of the $q^\text{th}$ bit labelling the $x_{k}$, i.e. $d_{Q(k-1)+q}$, with $q=1,\dots,Q$. 

We consider an equivalent baseband frequency selective channel with the impulse response $\mathbf{h}=[h_{L-1}, h_{L-2} \dots h_{0}]$, of delay spread $L$. Thus the received samples are
\begin{equation}
y_{k} =  \sum_{l=0}^{L-1}{h_{l} x_{k-l}} + w_{k},
\end{equation}
for $k=1,\dots,K$, where the noise $w_k$ is modelled as an additive white Gaussian noise (AWGN), with $\mathcal{CN}(0, \sigma_w^2)$, i.e. a zero mean Gaussian process with variance  $\sigma_w^2$.

The receiver is assumed to be ideally synchronized in time and frequency, and perfect channel state information is available.
We consider an iterative BICM receiver where a SISO channel decoder and a SISO symbol receiver exchange extrinsic information for iterative detection and decoding, as shown in Fig.~\ref{fig_0_iter_recv_model}.
A priori, extrinsic and a posteriori log likelihood ratios (LLRs) of coded bits $\mathbf{d}$ are respectively denoted $L_p(\cdot)$, $L_e(\cdot)$ and $L(\cdot)$, with respect to the SISO receiver.
The considered SISO symbol detector consists of a SISO channel equalizer and a symbol-wise SISO demapper module, as shown in Fig. \ref{fig_0_iter_recv_model}.  
The latter consits in a soft-output maximum a posteriori (MAP) demapper, and a soft-mapping unit, an APP distribution estimator, and the eventual use of the so-called ``Gaussian division'' operation for computing extrinsic symbol feedback (see discussion below and \cite{sahinCipriano_2018_DFEEP}).

The SISO equalizer computes an estimate $x^e_k$ of $x_k$, affected by a residual noise of variance $v^e_{x,k}$, whereas the SISO demapper uses these estimates to compute $L_e(\mathbf{d})$, and to deliver soft feedback $x^d_k$ to the equalizer for additional interference cancellation, such that $v^d_{x,k}$ is the variance of residual interference and noise of the feedback (this will be discussed in more detail afterwards).

In short, soft mapper uses LLRs from the decoder to estimate a prior distribution on $x_k=\alpha$, $\forall \alpha\in\mathcal{X}$
\begin{equation}
    \mathcal{P}_{k}(\alpha) \propto \textstyle\prod_{q=1}^{Q}e^{-\varphi^{-1}_q(\alpha)L_p(d_{k,q})}. \label{eq_priors}
\end{equation}
Soft demapper estimates a posteriori symbol distribution
\begin{equation}
    \mathcal{D}_{k}(\alpha) \propto \exp{\left(-\frac{|\alpha-x^e_{k}|^2}{v_{x,k}^e}\right)}\mathcal{P}_{k}(\alpha),\, \forall \alpha\in\mathcal{X}, 
    \label{eq_posteriors}
\end{equation}
which allows computing extrinsic LLRs towards the decoder
\begin{equation}
L_e(d_{k,j}) = \ln \frac{\sum_{\alpha\in\mathcal{X}_j^0}{\mathcal{D}_k(\alpha)}}{ \sum_{\alpha\in\mathcal{X}_j^1}{\mathcal{D}_k(\alpha)}} - L_p(d_{k,j}), \label{eq_demap_llr}
\end{equation}
with $\mathcal{X}_j^b=\lbrace \alpha\in\mathcal{X}: \varphi^{-1}_j(x)=b\rbrace$ where $b\in\mathbb{F}_2$.

\subsection{On SISO FIR DFE Structures and Problem Statement}
FIR structures are modelled with windowed processes; applying a sliding window $[-N_p,N_d]$ on $y_k$, we define $\mathbf{y}_k=[y_{k-N_p},\dots,y_{k+N_d}]^T$.
$N_p$ and $N_d$ are respectively the number of pre-cursor and post-cursor samples, and we denote $N\triangleq N_p+N_d+1$, and $N_p'\triangleq N_p+L-1$.  
Then, using the same window on $w_k$, and $[-N_p',N_d]$ on $x_k$, we have
\begin{equation}
\mathbf{y}_k = \mathbf{H}\mathbf{x}_k+\mathbf{w}_k, \label{eq_slidingwindow}
\end{equation}
with $\mathbf{H}$ being the $N\times N+L-1$ Toeplitz matrix generated by the static channel $\mathbf{h}$ with the first row being $\left[\mathbf{h}, \mathbf{0}_{1,N-1}\right]$.

Exact TV DFE receivers carry out interference cancellation with anti-causal estimates $\mathbf{\bar{x}}_k^{\mathbf{a}} \triangleq [\bar{x}^a_{k},\dots,\bar{x}^a_{k+N_d}]$ and causal estimates $\mathbf{\bar{x}}_k^{\mathbf{c}} \triangleq [\bar{x}^c_{k-N_p'},\dots,\bar{x}^c_{k-1}]$, with related variances $\mathbf{\bar{v}}_{\mathbf{x},k}^{\text{dfe}}\triangleq [ \bar{v}^c_{x,k-N_p'},\dots,\bar{v}^c_{x,k-1},\bar{v}^a_{x,k},\dots,\bar{v}^a_{x,k+N_d}]$. 
Causal estimates are directly dependent on the values of $(x_k^e, v_{x,k}^e)$ from previous turbo iterations, and they depend on mapping constraints. 
The equalized estimates are \cite{sahinCipriano_2018_DFEEP}
\begin{equation}
    \begin{array}{l}
        x^{e}_k = \bar{x}^{a}_k + \mathbf{f}_k^{}{}^H\mathbf{y}_k\\ 
        {}\qquad - \mathbf{g}^{\mathbf{c}}_k{}^H\mathbf{\bar{x}}_k^{\mathbf{c}} - \mathbf{g}^{\mathbf{a}}_k{}^H\mathbf{\bar{x}}_k^{\mathbf{a}}\\
        v^{e}_{x,k} = 1/\xi_k  - \bar{v}^{a}_{x,k}
    \end{array}
    ,\, \text{ } 
    \begin{cases}
        \mathbf{f}_k \triangleq \mathbf{\Sigma}_k{}^{-1}\mathbf{h}_0/\xi_k, \\
        \xi_k \triangleq \mathbf{h}_0^H\mathbf{\Sigma}_k^{-1}\mathbf{h}_0,
    \end{cases}
    \label{eq_dfetv_model}
\end{equation}
with 
$\mathbf{\Sigma}_k \triangleq \sigma_w^2\mathbf{I}_N + \mathbf{H}\mathbf{Diag}(\mathbf{\bar{v}}^\text{dfe}_k)\mathbf{H}^H$, $\mathbf{g}^{\mathbf{c}}_k \triangleq [\mathbf{H}^H\mathbf{f}_k]_{1:N_p'}$, $\mathbf{g}^{\mathbf{a}}_k \triangleq [\mathbf{H}^H\mathbf{f}_k]_{N_p'+1:N_p'+N_d}$ and $\mathbf{h}_0 \triangleq [\mathbf{H}]_{:,N_p'+1}$. 
In this paper, for numerical results the values of window parameters are taken as $N=3L+2$ and $N_d=2L$.

\begin{figure*}[!t]
	\centering
    
	\includegraphics[width=0.55\textwidth]{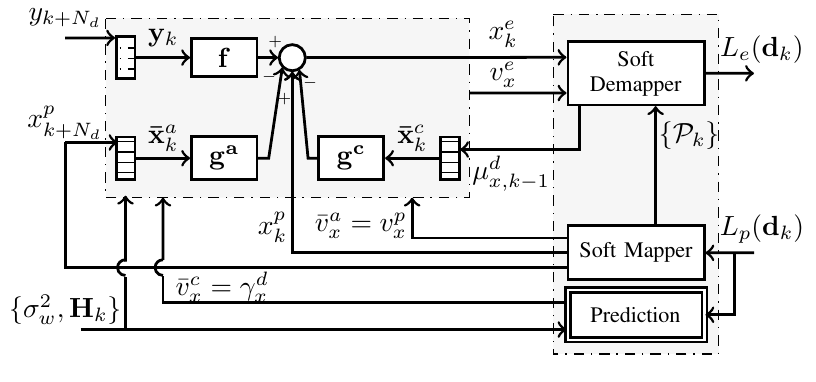}
	\caption{Soft-Input Soft-Output IV DFE APP structure.}
	\label{fig_1_IVDFEIC_APP}
\end{figure*}

IV DFE is obtained when $\mathbf{\bar{v}}^\text{dfe}_k$ is independent of $k$, with $\mathbf{\bar{v}}^\text{dfe}_k=\mathbf{\bar{v}}^\text{dfe}$, all filters being invariant,  $\mathbf{f}$, $\mathbf{g}^{\mathbf{c}}$ and $\mathbf{g}^{\textbf{a}}$, as in \cite{tao_2016_low}:
\begin{equation}
    \begin{array}{l}
        x^{e}_k = \bar{x}^{a}_k + \mathbf{f}^{}{}^H\mathbf{y}_k\\ 
        {}\qquad - \mathbf{g}^{\mathbf{c}}{}^H\mathbf{\bar{x}}_k^{\mathbf{c}} - \mathbf{g}^{\mathbf{a}}{}^H\mathbf{\bar{x}}_k^{\mathbf{a}}\\
        v^{e}_x = 1/\xi  - \bar{v}^{a}_x
    \end{array}
    ,\, \text{ } 
    \begin{cases}
        \mathbf{f} \triangleq \mathbf{\Sigma}{}^{-1}\mathbf{h}_0/\xi, \\
        \xi \triangleq \mathbf{h}_0^H\mathbf{\Sigma}^{-1}\mathbf{h}_0.
    \end{cases}
    \label{eq_dfeiv_model}
\end{equation}
The variances of soft interference cancellation estimates are
\begin{equation}
    \label{eq_iv_variance}
    \mathbf{\bar{v}}^\text{dfe}_\mathbf{x} = [\bar{v}^c_x \mathbf{1}_{N_p',1}, \bar{v}^a_x \mathbf{1}_{N_d+1,1}],
\end{equation}
where $\bar{v}^a_x$ and $\bar{v}^c_x$ are respectively the overall reliability of anti-causal and causal estimates.
For interference cancellation, the set of anti-causal estimates are available before equalization, and an accurate estimate of their reliability is given by the least-squares estimation;
    $\bar{v}^a_x = K^{-1} \textstyle \sum_{k=0}^{K-1} \bar{v}^a_{x,k}$.
In most SISO DFE structures, the anti-causal estimates are the prior estimates given by the decoder\footnote{However, note that in the self-iterated SISO DFE of \cite{sahinCipriano_2018_DFEEP}, the anti-causal estimates are the causal estimates of previous iterations.}; ${x}^p_k \triangleq\mathbb{E}_{\mathcal{P}_k}[x_k], v^p_{x,k} \triangleq\text{Var}_{\mathcal{P}_k}[x_k]$.

As stated in the introduction, the core of the problem lies in the computation of $\bar{v}^c_x$.
A simple, but inaccurate solution to this is the ``perfect decision assumption''~: $\bar{x}^c_k$ are all assumed to be equal to $x_k$, yielding $\bar{v}^c_x=0$. 
This approach is sufficient at high SNR operating points, but as shown in \cite{tuchler_turbo_2002,sahinCipriano_2018_DFEEP}, it degrades performance in moderately or severely selective channels. 

\begin{figure*}[!t]
	\centering
    
	\includegraphics[width=0.65\textwidth]{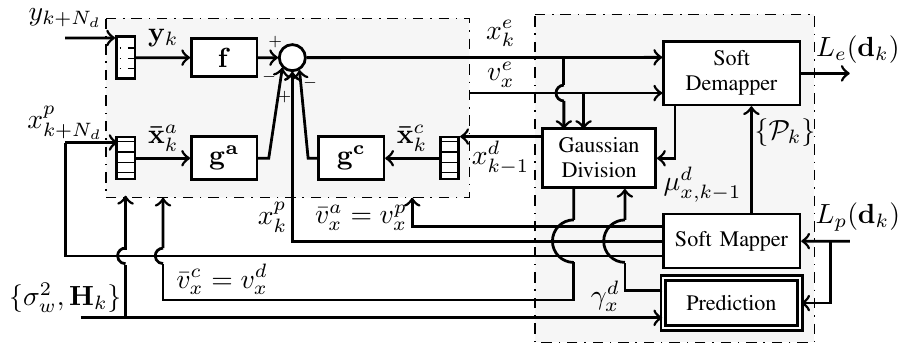}
	\caption{Soft-Input Soft-Output IV DFE EP structure.}
	\label{fig_2_IVDFEIC_EP}
\end{figure*}

Hence, with these notations, this paper's main objective is to evaluate novel prediction methods to compute $\bar{x}^c_k$ for optimizing the inter-symbol interference (ISI) mitigation performance of IV DFE.
In the following the cases of IV DFE based on APP-based soft feedback (see Fig.~\ref{fig_1_IVDFEIC_APP}), and EP-based soft feedback (see Fig.~\ref{fig_2_IVDFEIC_EP}) will be discussed.

\subsection{APP Soft Feedback Computation}\label{ssec:app_fb}

A common approach to compute soft feedback for DFE is to use APP estimates, as in IV structures of \cite{lopes_soft-feedback_2006,lou_soft_2014,tao_2016_low}. Derivation of a DFE APP for TV FIR is available in \cite{sahinCipriano_2018_DFEEP}.

These soft estimates are given by the mean and the variance of the posterior symbol distribution $\mathcal{D}_{k}$, given in Eq.~(\ref{eq_posteriors}). We denote 
\begin{equation}
    \label{eq_dem_stats}
    \begin{aligned}
        \mu^d_k&\triangleq\mathbb{E}_{\mathcal{D}_k}[x_k]&={}&\textstyle\sum_{\alpha\in\mathcal{X}}\alpha \mathcal{D}_k(\alpha),\\
        \gamma^d_{x,k}&\triangleq\text{Var}_{\mathcal{D}_k}[x_k]&={}&\textstyle\sum_{\alpha\in\mathcal{X}}\vert\alpha\vert^2 \mathcal{D}_k(\alpha) - \vert \mu^d_k \vert^2.
    \end{aligned}
\end{equation}
For the IV DFE APP filter computation, an invariant variance $\gamma^d_x$ is needed, as with the causal reliability $\bar{v}^c_x$. 
Unlike the anti-causal reliability, $\gamma^d_x$ cannot be estimated using the causal estimates $(\bar{x}_k^c,\bar{v}^c_{x,k})=(\mu_k^d,\gamma_{x,k}^d)$, as causal estimates are only available once the filter is computed, and the equalization is being carried out. 
Thus, a predictive estimation is required.
This receiver corresponds to the Fig.~\ref{fig_1_IVDFEIC_APP}, with $\bar{v}^c_x=\gamma^d_x$.

\subsection{Predictive EP-based Soft Feedback Computation}\label{ssec:ep_fb}

In \cite{sahinCipriano_2018_DFEEP}, an expectation propagation (EP) based soft feedback is used within a TV DFE which proved to bring several performance improvements.
Unlike APP estimates, EP-based estimates carry only the extrinsic information brought by the demapper, and prevents DFE from relying on its own bias, and improves the asymptotic predictability of the receiver. 
For TV DFE, these estimates are obtained by the division of two Gaussian PDFs, which yields another Gaussian PDF with mean and variance given by
\begin{equation}
    {x}^d_k = \frac{\mu^d_k v^e_x - x^e_k\gamma^d_{x,k}} {v^e_x-\gamma^d_{x,k}},\, \text{and},\, {v}^d_{x,k} = \frac{ v^e_x \gamma^d_{x,k}} {v^e_x-\gamma^d_{x,k}}.
\end{equation}
For the sake of simplicity, this operation is referred to as ``Gaussian division''.

For IV DFE EP structure, this feedback is not adapted, as the invariant filter is unable to adapt its coefficients to handle the strong variations of $v^d_{x,k}$, which depends on the instantaneous APP variance $\gamma_{x,k}^d$. 
Hence we propose to use a feedback based on overall APP reliability, with
\begin{equation}
    {x}^d_k = \frac{\mu^d_k v^e_x - x^e_k\gamma^d_x} {v^e_x-\gamma^d_x},\, \text{and},\, {v}^d_{x,k} = \frac{ v^e_x \gamma^d_x} {v^e_x-\gamma^d_x}, \label{eq_demap_ep_stats}
\end{equation}
where a predicted invariant APP variance $\gamma^d_x$ is used to generate the feedback. 
Moreover EP-based estimates have an invariant variance, i.e. $v^d_x \triangleq v^d_{x,k}, \forall k$, as the causal reliability is directly related to the predicted APP variance.
The receiver based on this structure is illustrated in the Fig.~\ref{fig_2_IVDFEIC_EP}, with $\bar{v}^c_x=v^d_x$.

\section{Semi-Analytic Abstraction of SISO FIR DFE}\label{sec:abstraction}

In this section, a prediction model for the turbo DFE structure of Equation~(\ref{eq_dfeiv_model}) is exposed, without loss of generality, for the case where anti-causal estimates are given by decoder's extrinsic LLRs, i.e. $\bar{v}^a_x = v^p_x$.
Such models are usually used for handling physical layer link quality prediction that is necessary to enable link adaptation with low computational complexity. The originality is that, in our context, it will be exploited for online estimation of the reliability of causal estimates for SISO DFE filter computation.

\subsection{General Structure and Analytical Equalizer Model}

SISO DFE is modelled with two independent components; an analytical model for the equalizer, and a numerical model for the soft demapper. 
Unlike asymptotic transfer models ($K \rightarrow + \infty$) used in extrinsic information transfer (EXIT) analysis \cite{brink_2000_designing}, finite-length transfer models are used for characterizing the demapper, as prior works on performance prediction noted their positive impact on accuracy \cite{sabbaghian_2009_analyticalapproach, visoz_semi-analytical_2010}.

\newpage

Following Eq.~(\ref{eq_dfeiv_model}), the IV DFE-IC output reliability $v^e_x$ is modelled by a function $\phi_\text{REC}$ as
\begin{equation}
    \label{eq_rec_model}
    \begin{split}
    v^e_x &= \phi_\text{REC}(\sigma_w^2, \mathbf{h}, v^p_x , \bar{v}^c_x)\\
    &\triangleq \left(\mathbf{h}_0^H \left[ \sigma_w^2\mathbf{I}_N + \mathbf{H}\mathbf{Diag}(\mathbf{\bar{v}}^\text{dfe}_\mathbf{x})\mathbf{H}^H \right]^{-1}\mathbf{h}_0 \right)^{-1}  - v^p_x,
    \end{split}
\end{equation}
where $\mathbf{\bar{v}}^\text{dfe}_\mathbf{x}$ is given by Eq.~(\ref{eq_iv_variance}), with $\bar{v}^a_x = v^p_x$. 
This function is strictly increasing with $\bar{v}^c_x \in [0, \sigma_x^2]$.

As an analytical model is unavailable for characterizing the demapper, it is modelled with a look-up table (LUT) $\phi_\text{DEM}$, given by
\begin{equation}
    \bar{v}^c_x = \phi_\text{DEM}(v^e_x, \cdot),    \label{eq_dem_model}
\end{equation}
where $\bar{v}^c_x$ is the expected value of causal estimates' variance, taken over realizations of the channel noise, the equalizer outputs and the prior LLRs. 
The second argument `$\cdot$' in Equation~(\ref{eq_dem_model}) models prior information, and the exact nature of the argument depends on the selected prediction approach.
Improvements proposed in the upcoming subsection concern this module.

Since equalizer and demapper iteratively exchange reliabilities, the two functions representing their model must be composed to yield a recursive equation on $\bar{v}^c_x$. 
Hence by using $n=0,\dots,N_\text{pred}$ for indexing recursions, we have 
\begin{equation}
    \bar{v}^c_x[n+1] = \phi_\text{DEM}(\phi_\text{REC}(\sigma_w^2, \mathbf{h}, v^p_x, \bar{v}^c_x[n]), \cdot) \triangleq f_\text{pred}(\bar{v}^c_x[n]). \label{eq_fixedpoint}
\end{equation}
If $f_\text{pred}$ admits a unique fixed-point on $\bar{v}^c_x$, then 
the desired predicted reliability estimate is this fixed-point. Moreover,
the optimality of IV DFE-IC strongly depends on $\bar{v}^c_x$ and hence on the accuracy of $\Phi_\text{DEM}$.

\begin{figure}[!t]
	\centering

	\includegraphics[width=0.48\textwidth]{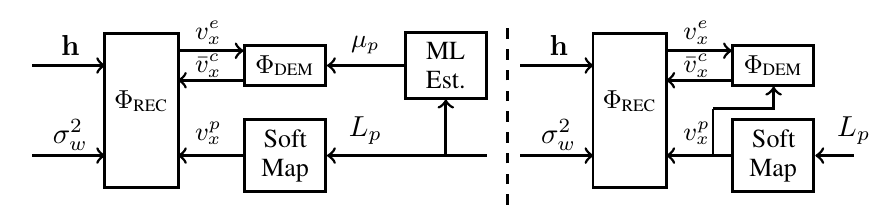}  

    \caption{Block diagrams for bit-wise (left) and symbol-wise (right) causal reliability prediction schemes.}
	\label{fig_3_strct_pred}
\end{figure}

Fig.~\ref{fig_3_strct_pred} illustrates reliability prediction structures with the two semi-analytical models that will be introduced below.

\subsection{Numerical Demapping Models for APP/EP}

Modelling the demapper with prior information is challenging due to its highly non-linear behavior, and due to strong simplifying assumptions.
The main focus of the numerical model will be the model of the posterior symbol distribution's  variance $\gamma^d_x$, required for using the DFE with APP feedback, discussed in Subsection~\ref{ssec:app_fb}.
Besides, as the variance $v^d_x$ of the proposed predictive EP feedback, in Subsection~\ref{ssec:ep_fb}, is analytically linked to $\gamma^d_x$, numerical model of APP estimates is common for both types of feedback.

\subsubsection{Mutual Information based Prediction (Bit-wise)}

In the BPSK receiver of \cite{lopes_soft-feedback_2006}, a prediction scheme is considered, assuming input/output LLRs of the demapper to be consistent Gaussian, i.e. $L_{(\cdot)}(d_{k,q})\sim \mathcal{N}(\bar{d}_{k,q}\mu_{(\cdot)}, 2\mu_{(\cdot)})$, where $\bar{d}_{k,q}=1-2d_{k,q}$, and where $(\cdot)$ is $p, e$ or void, depending on concerned LLRs. 
Using a semi-analytical density evolution, parameter $\mu_e$ of extrinsic LLRs is predicted using $\mu_p$. 
The parameter $\mu_{(\cdot)}$ is bijectively linked to the average mutual information (MI) between LLRs and the associated coded bits, usable for binary prediction as in \cite{visoz_semi-analytical_2010}. 
Hence using such formalism, the approach of \cite{lopes_soft-feedback_2006} can be extended to any constellation and mapping.

More specifically, the demapper behaviour is numerically integrated for each $\gamma_{x,k}^d$, $k=1,\dots,K$, under the assumption that prior LLRs are consistent Gaussian with the parameter $\mu_p$, and the assumption that the residual ISI and noise affecting the equalized symbols $x_k^e$ are Gaussian-distributed, i.e. $x_k^e \sim \mathcal{CN}(x_k, v^e_x)$.
Hence with these conditions, a LUT on $\mu_p$ and $v^e_x$ is built with 
\begin{align}
    \bar{v}^c_x = \phi_\text{DEM}(v^e_x, \mu_p) \triangleq \frac{1}{K} \sum_{k=1}^{K} \mathbb{E}_{\mathbf{L}_\mathbf{p}, x^e}[\bar{v}^c_{x,k}], \label{eq_feedback_expectation}
\end{align}
where $\bar{v}^c_x$ is the variance of residual error on APP/EP soft symbols and the priors' parameter $\mu_p$ is measured with a maximum-likelihood (ML) estimator (see Fig. \ref{fig_3_strct_pred}, left)
\begin{equation}
    \mu_p \approx \sqrt{1+\sum_{k=1}^{K}\sum_{q=1}^{Q}\lvert L_p(d_{k,q})\rvert^2}-1.
\end{equation}

In the case of APP feedback, i.e. for $\bar{v}^c_{x,k}=\gamma_{x,k}^d$, the expectation in Equation~(\ref{eq_feedback_expectation}) becomes
\begin{align}
    &\textstyle \mathbb{E}_{\mathbf{L}_\mathbf{p}, x^e}[\gamma_{x,k}^d] =  \frac{1}{M}\sum_{x_k \in \mathcal{X}} \text{Var}\left[\mathcal{D}(x_k,x^e_k,\mathbf{L}_{\mathbf{p},k})\right] \nonumber\\
    &\textstyle \quad \mathcal{CN}(x_k^e;x_k, v^e_x) \prod_{q=1}^Q \mathcal{N}(L_{p,k,q};\bar{\varphi}^{-1}_q(x_k)\mu_p,2\mu_p)
\end{align}
where the APP probability mass function of a dummy symbol $x\in\mathcal{X}$ is
\begin{equation}
    \mathcal{D}(x,x^e,\mathbf{L}_{\mathbf{p}}) \triangleq \frac{1}{Z} \exp{\left(-\frac{|x-x^e|^2}{v^e} - \sum_{q=1}^{Q}\varphi^{-1}_q(x^e)L_{p,q} \right)}, 
\end{equation}
where $Z$ is the normalization constant such that $\sum_{x \in \mathcal{X}} \mathcal{D}(x,x^e,\mathbf{L}_{\mathbf{p}})=1$.
Moreover, considering the EP feedback's analytical expression in Equation~(\ref{eq_demap_ep_stats}), we have 
\begin{equation}
    \mathbb{E}_{\mathbf{L}_\mathbf{p}, x^e}[\gamma_{x,k}^d] = \left( \left( \frac{1}{K} \sum_{k=1}^{K}\mathbb{E}_{\mathbf{L}_\mathbf{p}, x^e}[\gamma_{x,k}^d] \right)^{-1} - \frac{1}{v^e_x}\right)^{-1}.
\end{equation}

The binary prediction scheme above appeared to yield too optimistic estimates in \cite{lopes_soft-feedback_2006}, and instead \emph{Lopes et al.} resorted to obtain $\mu_e$ and $\mu_p$ through BPSK channel estimators, which circumvents consistent Gaussian LLR approximation.

More specifically, this problem ensues from well known issues with regards to performance prediction of turbo iterative systems, for which the consistent Gaussian approximation of LLRs was shown to be only accurate at the zeroth turbo-iteration, and in the asymptotic limit.
Otherwise inaccurate estimates propagate across turbo iterations due to the internal non-linear dynamics of channel decoding \cite{fu_2005_stochasticanal}. 
To overcome this prediction bias, prediction based on a two-parameter LLRs' model has been shown to be much more accurate \cite{ibing_2010_predictingconvergence}. 
Such models consider $L_{(\cdot)}(d_{k,q})\sim \mathcal{N}(\bar{d}_{k,q}\mu_{(\cdot)}, \eta_{(\cdot)}\mu_{(\cdot)})$, where $\eta_{(\cdot)}$ is no longer 2.
The ML estimator used for measuring $\mu_p$ in the binary prediction is very sensitive to $\eta_{p}$, which is the reason why the binary prediction is not robust enough in practice.

\begin{figure}[t]
	\centering

	\includegraphics[width=0.48\textwidth]{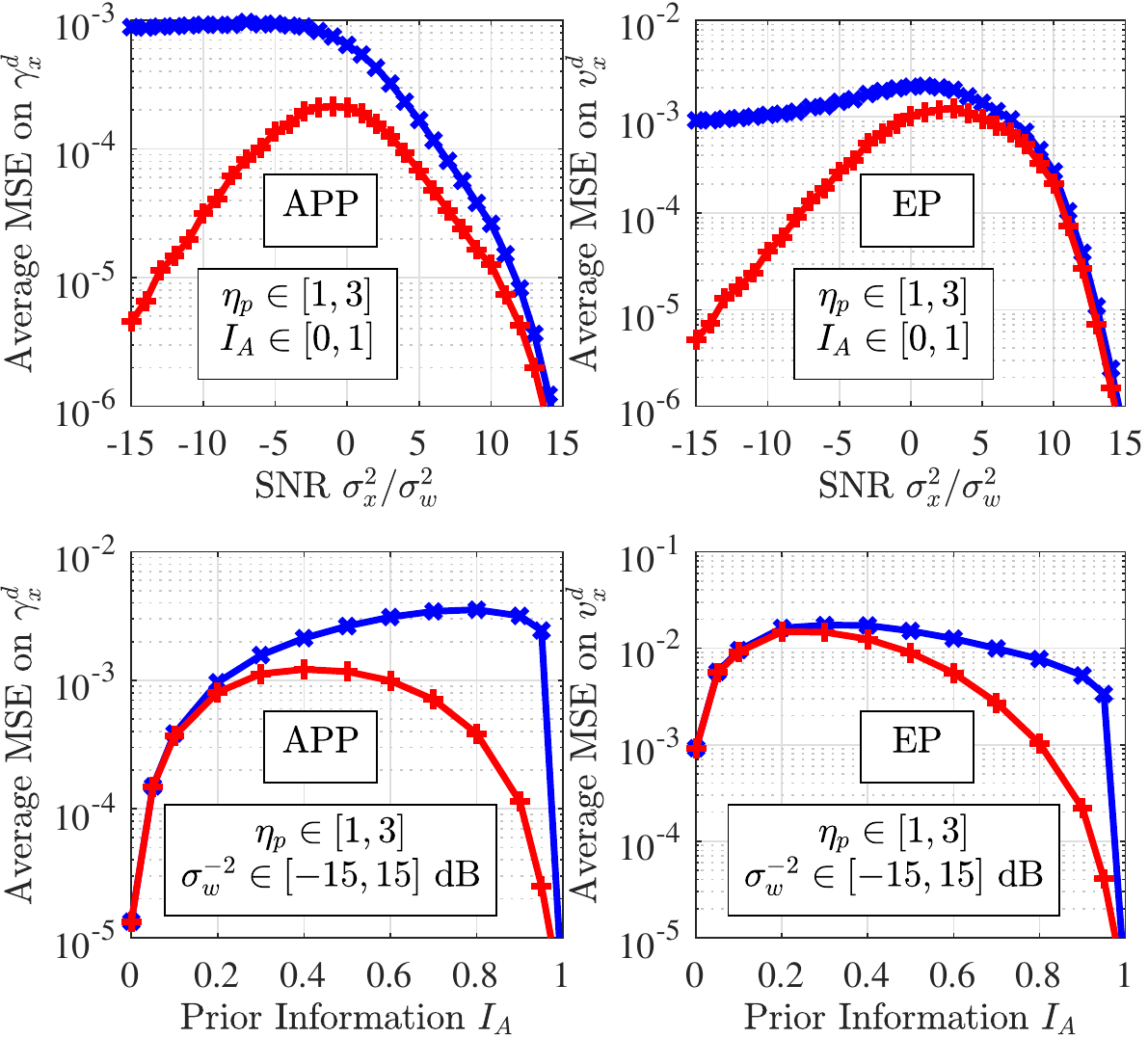}

    \caption{Mean-square error on the prediction quality of the binary (blue, $\times$) and symbol-wise (red, +) schemes.}
	\label{fig_4_pred_quality}
\end{figure}

\begin{figure}[t]
	\centering

    \includegraphics[width=0.45\textwidth]{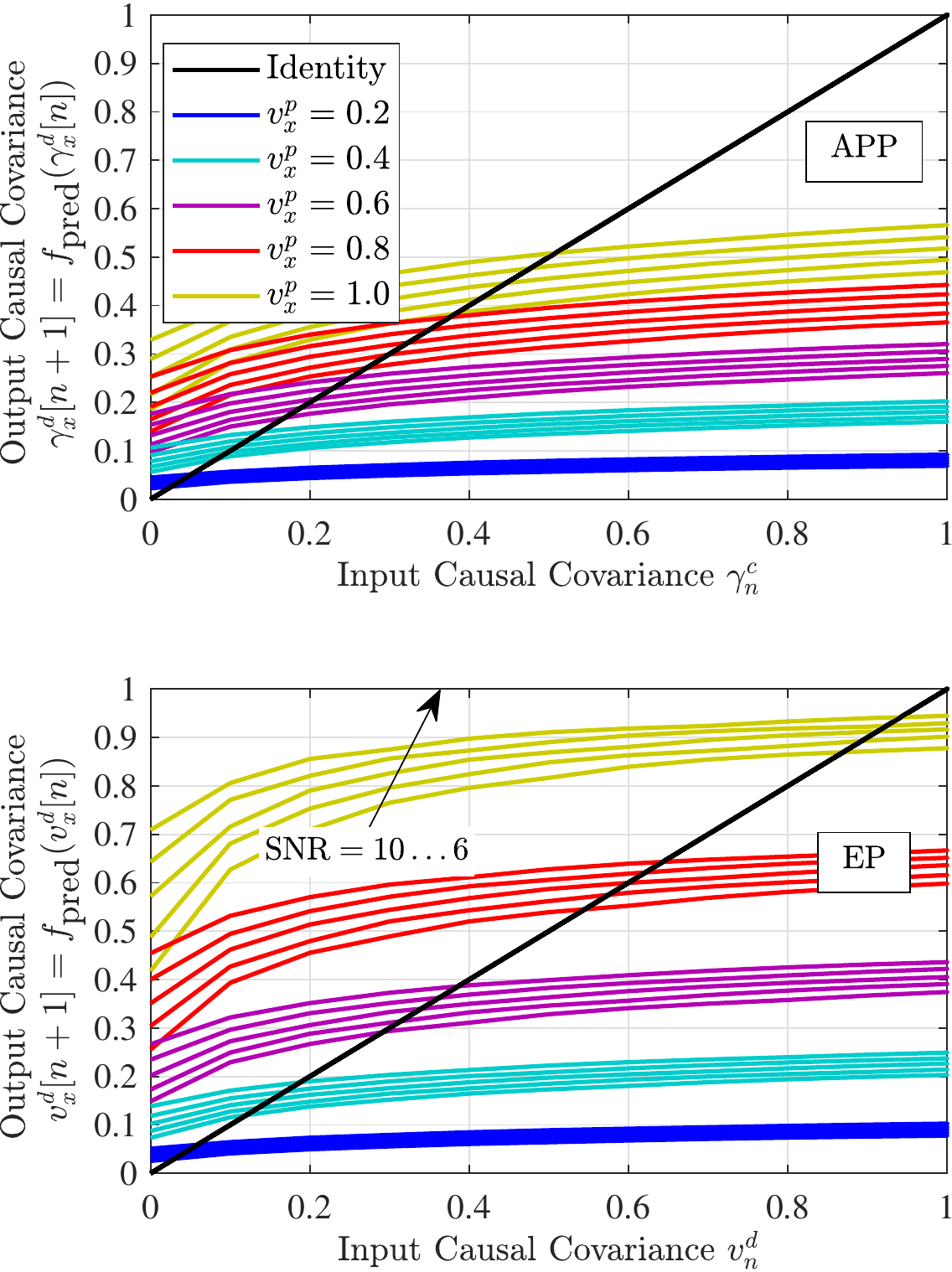}
	
	\caption{Fixed-points of the symbol-wise $f_\text{pred}$ for SNR varying from 6 to 10 dB, for each value of prior reliability.}
	\label{fig_5_pred_fixed_pt}
\end{figure}

\subsubsection{Prior Variance based Prediction (Symbol-wise)}
For our context, two-parameter models are too complex as they require expensive online parameter estimators to get both $\mu_p$ and $\eta_p$.
Hence, a single-parameter demapper model with reasonable estimation complexity has been preferred. 
We searched for the parameter which is the least sensitive to the variations of prior LLRs' variance-to-mean ratio $\eta_{p}$. 

Following a thorough and almost exhaustive study of the different alternative parameters for tracking evolution of $\bar{v}^c_x$, anti-causal variance $v^p_x$ has been found to be sensitive to the changes on $\eta_p$, very similarly to $\bar{v}^c_x$, with the advantage of $v^p_x$ being directly computable online using a simple least-squares estimation. Hence, we propose the following LUT
\begin{equation}
    \bar{v}^c_x = \phi_\text{DEM}(v^e_x, v^p_x),\,\textstyle \text{ } 
    \begin{cases}
        \bar{v}^c_x  \triangleq K^{-1} \sum_{k=1}^{K} \mathbb{E}_{\mathbf{L}_\mathbf{p}, x^e}[\bar{v}^c_{x,k}], \\
        v^p_x \triangleq K^{-1} \sum_{k=1}^{K} \mathbb{E}_{\mathbf{L}_\mathbf{p}}[v^p_{x,k}],
    \end{cases}
\end{equation}
where both input $v^p_x$ and output $\bar{v}^c_x$ are numerically integrated using prior LLRs generated for a fixed value of $\eta_p$. 
Indeed, as $\eta_p$ cannot be accurately measured online, the conventional consistent approximation \cite{visoz_semi-analytical_2010} is kept with $\eta_p=2$. 
In the following, we will assess its impact on the prediction accuracy.

\subsubsection{Robustness of demapper prediction}

The sensitivity of the considered prediction schemes to variations in $\eta_{p}$ is evaluated. This aspect is important for characterizing the robustness of iterative receiver prediction schemes, as the hypothesis $\eta_{p}=2$, used for LUT generation, is only true at the initial turbo-iteration and then it varies \cite{fu_2005_stochasticanal}.

\begin{figure*}[t]
	\centering
    
    \includegraphics[width=0.695\textwidth]{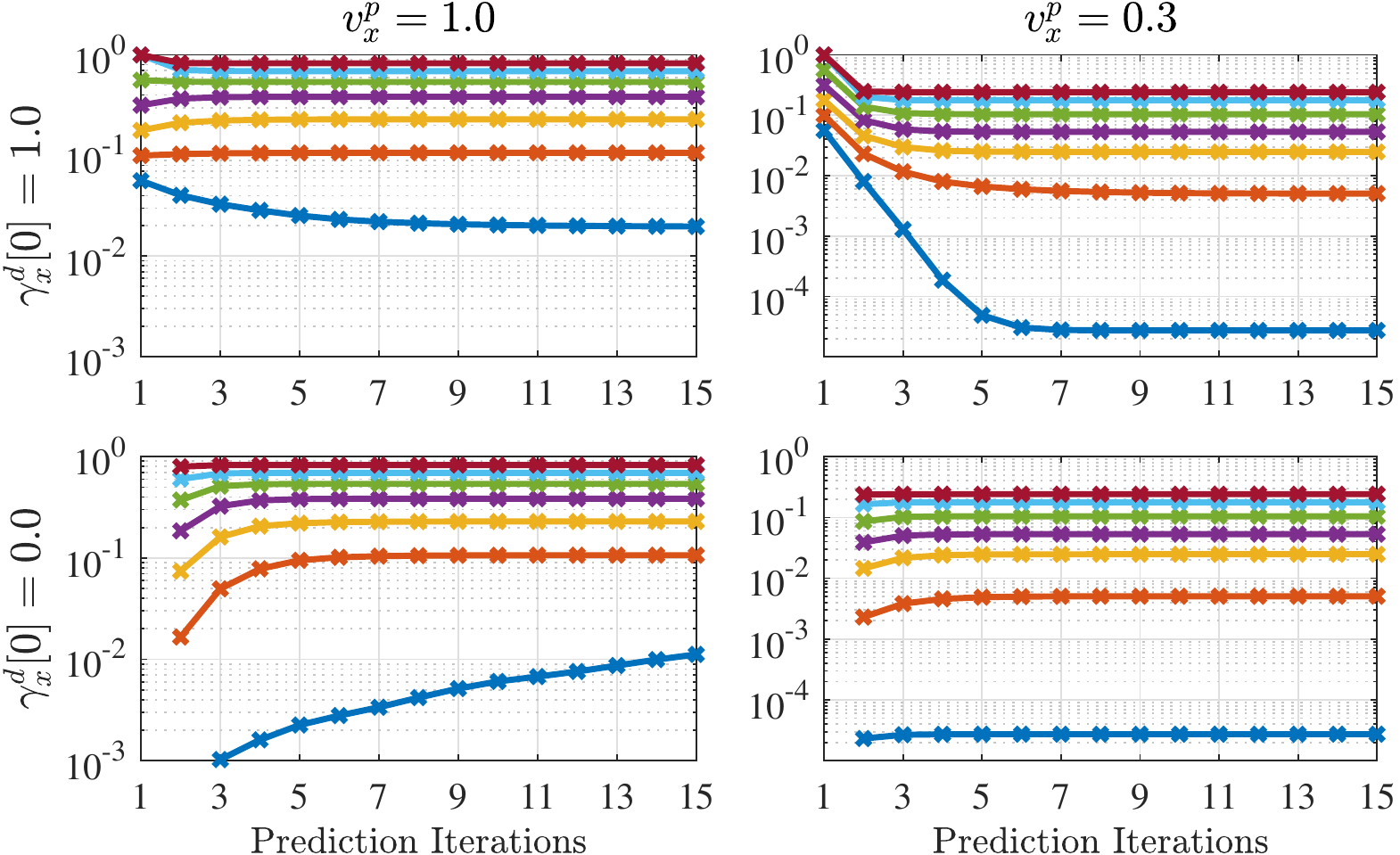}
	
	\caption{Evolution of APP covariance prediction $\gamma^d_{n+1} = f_\text{pred}(\gamma^d_{n})$. Different colors of plots are for SNR varying from -5 to 25 dB, with 5 dB steps (towards lower plots).}
	\label{fig_6_pred_conv_speed}
\end{figure*}

An AWGN channel is simulated with blocks of 16-QAM symbols with $K=1024$, to emulate the output $x^e$ of the equalizer, for $v^e$ varying from $-15$ to $15$~dB, along with Gaussian-distributed prior LLRs generated with prior MI $I_A$ varying from 0 to 1 bit (and hence determining $\mu_p$), with $\eta_p$ varying from 1 to 3. The average mean squared error (MSE) between the predicted causal covariance and true causal covariance is measured, and plotted in Fig.~\ref{fig_4_pred_quality}. 
The left side of the figure provides results for APP feedback, and the right side for EP-based feedback.
The binary approach is seen to be severely impacted by the changes in $\eta_p$, whereas the symbol-wise approach, although not perfect, remains more robust. Considerable differences are seen at low to medium SNR for high prior information, which suggests that symbol-wise schemes would have an advantage at the decoding threshold in asymptotic behaviour, i.e. when a high number of turbo-iterations are used. Oppositely, without any turbo-iteration, both schemes would perform identically.

\subsection{Convergence Analysis}

The convergence of the proposed iterative semi-analytical prediction schemes could be assessed formally through fixed-point analysis of Eq. (\ref{eq_fixedpoint}).
However, due to the untractable non-linear expression of  $\Phi_\text{DEM}$, an analytic approach is not possible, and we resort to numerical evaluations. 
In the following, the convergence of the symbol-wise prediction scheme is evaluated.

Numerical evaluations of the proposed $f_\text{pred}$ show that we can reasonably conjecture that this function is continuous on the interval $[0,+\infty[$, with a Lipschitz constant strictly less than one, for all $\sigma_w^2 \geq 0$ and $0 \leq v^p_x \leq \sigma_x^2$. This ensures that Eq. (\ref{eq_fixedpoint}) reaches a unique fixed-point $\bar{v}^c_x \in [0,+\infty[$ for any initial guess. 
This conjecture has been checked for various channels $\mathbf{h}$, and Fig.~\ref{fig_5_pred_fixed_pt} plots $f_\text{pred}$ for the Proakis-C channel ($\mathbf{h}=[1,2,3,2,1]/\sqrt{19}$), using the symbol-wise demapper model for the Gray-mapped 16-QAM constellation.

The convergence speed of the prediction scheme is also evaluated numerically in order to determine optimal parameters of the algorithm.
Indeed, the fixed-point $\bar{v}^c_x=\bar{v}^c_x[\infty]$ is reached more or less quickly depending on if the initial value $\bar{v}^c_x[0]$ is close to $\bar{v}^c_x[\infty]$. 

In particular, due to the near flat evolution of $f_\text{pred}$ for $\bar{v}^c_x$ close to $\sigma_x^2=1$, initializing with $\bar{v}^c_x[0]=1$ results in fast convergence at low SNRs, and high anti-causal covariance, but slow convergence otherwise.
Oppositely with  $\bar{v}^c_x[0]=0$ faster convergence is achieved for high SNRs and low anti-causal covariance.
This behaviour is illustrated for Proakis-C 16-QAM APP covariance in Figure~\ref{fig_6_pred_conv_speed}.

We propose to use the heuristic $\bar{v}^c_x[0]= \min(1, \sigma_w)$, when $v^p_x>0.5$, where the standard deviation of the channel noise is experimentally shown to serve as a convergence accelerating heuristic, when the channel is normalized. 
Otherwise using  $\bar{v}^c_x[0]=0$ is preferable for faster convergence.

\begin{figure*}[t]
	\centering

    \includegraphics[width=\textwidth]{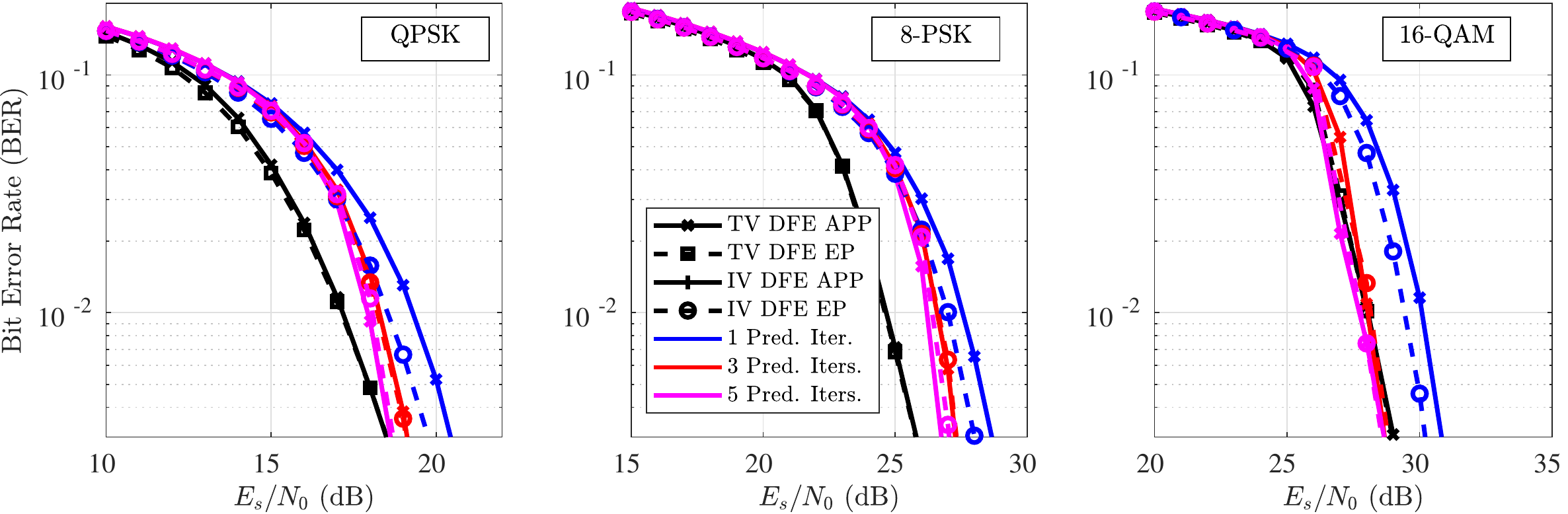}
	\caption{Uncoded bit-error rate (BER) performance of proposed predictive IV DFEs.}
	\label{fig_ber_uncode}
\end{figure*}

\section{Numerical Results}\label{sec:perfos}
\subsection{Uncoded equalization behaviour}

In this paragraph, the uncoded finite-length behaviour of the proposed IV DFE with online prediction is evaluated.
Exact TV DFE counterparts are used as lower-bound references on bit-error rate (BER), to assess the prediction accuracy. Note IV FIR receivers might outperform TV FIRs in some cases \cite{jeongMoon_2013_selfiteratingSoftEqualizer}, as the latter are more sensitive to the convergence errors committed by the SISO decoder.

Block transmission in Proakis C channel is considered with $K=256$ and with QPSK, 8-PSK and 16-QAM constellations.
In Fig.~\ref{fig_ber_uncode}, BER of TV DFE with APP and EP feedback are compared to the proposed predictive IV implementations.
IV DFE converges towards the curve of TV counterparts, especially at high SNR, but it is seen that a gap remains at medium BER for some constellations, due to dynamic filtering capabilities of TV receivers. 
EP feedback is shown to be mostly equivalent to APP feedback in this uncoded use case, but at high BER, EP has an advantage over APP both for TV and IV receivers, which suggests that improved decoding thresholds can be obtained with channel coding.

\subsection{On the Operating Regions of FIR Receivers}

A previous work on TV FIR turbo equalizers concluded that TV DFE significantly outperforms TV LE at high data rates \cite{sahinCipriano_2018_DFEEP}, whereas TV LE remains preferable at very low rates, as it achieves same performance with less complexity. 
In the following, the asymptotic behaviour and the computational complexity of the proposed receiver along with IV FIRs is evaluated in a similar manner.

\begin{figure}[t]
	\centering

    \includegraphics[width=0.44\textwidth]{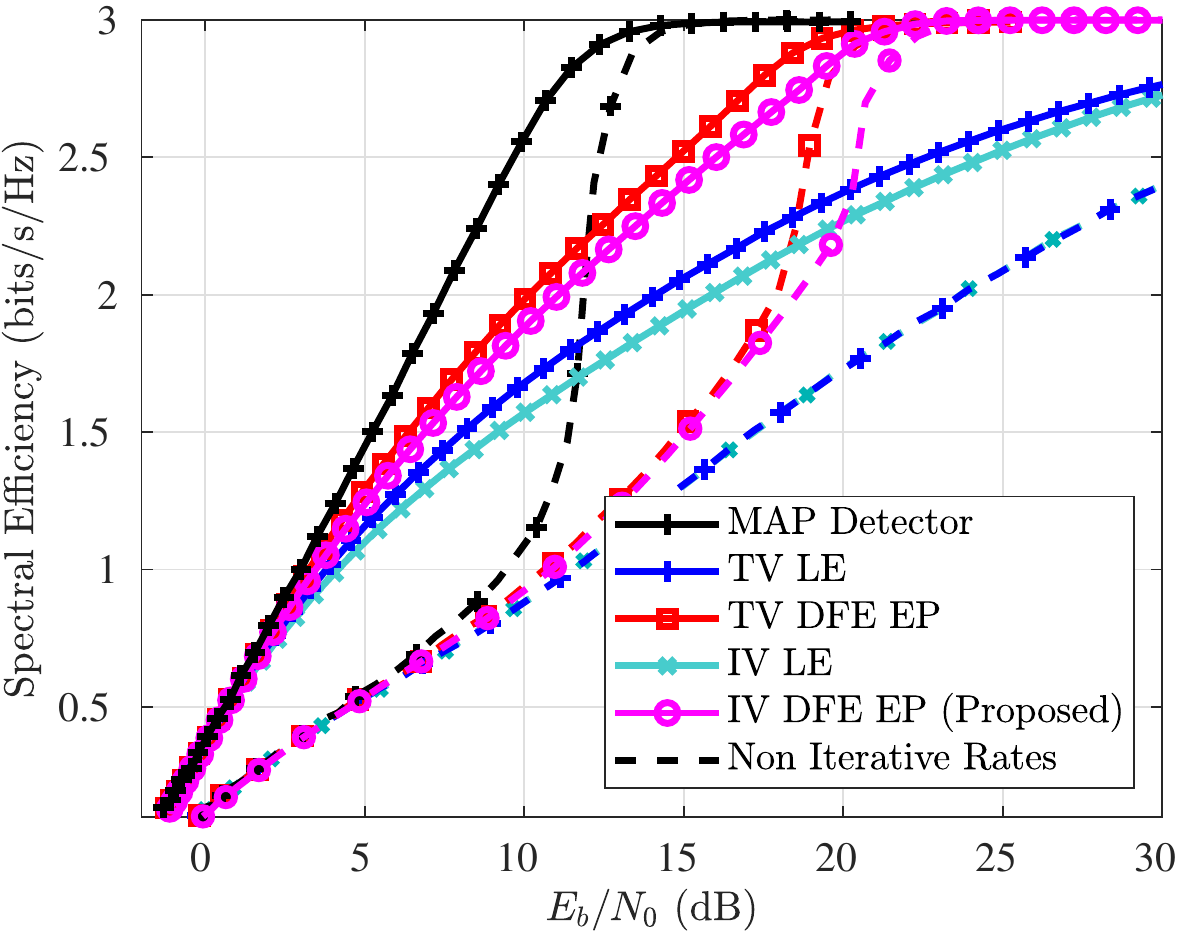}

	\caption{Achievable rates of FIR receivers for 8-PSK in Proakis C channel.
	}
	\label{fig_asymp}
\end{figure}

Through the extrinsic information transfer (EXIT) analysis of a SISO module, a mutual information (MI) based transfer function model, $I_E=\mathcal{T}_R(I_A, \mathbf{h}, \sigma_w^2)$ is obtained \cite{brink_2000_designing}, where $I_A$ and $I_E$ denote respectively the MI between coded bits and the prior LLRs and the extrinsic LLRs. 

EXIT functions notably allow to numerically predict the achievable rates of SISO receivers, through the area theorem of EXIT charts \cite{hagenauer_exit_2004}.
Indeed, MAP detector's EXIT chart's area yields an accurate prediction of the channel symmetric information rate (SIR) \cite{arnold_simulation-based_2006}, the highest possible transmission rate for practical constellations, without channel knowledge at the transmitter. 
However, for approximate receivers which violate the extrinsic message principle of turbo detection, the rates predicted by EXIT can be too optimistic (e.g. SIR of APP DFE appears to surpass MAP, which is impossible). 
This has been observed for APP-based receivers in \cite{sahinCipriano_2018_DFEEP,liWuTao_2019_perfAnalysisAndImprovemetForVAMPFDE}, but the EP-based DFE does not suffer from this phenomenon.
Hence in the following, the proposed predictive EP-based IV DFE is evaluated.

IV DFE EP with symbol-wise prediction scheme is used for 8-PSK transmissions in the Proakis C channel, and numerically obtained achievable rates are plotted in solid lines in Fig. \ref{fig_asymp}.
Dotted plots illustrate the achievable rates without turbo-iterations, for each receiver.
IV receivers are shown to follow the behaviour of their TV counterpart within a gap of about $0.1$~bits/s/Hz for both LE and DFE, but IV DFE still keep a significant upper hand over TV LE at medium and high spectral efficiency operating points.
Using IV FIR receivers to operate at a given rate requires about $1.5$~dB more energy than TV FIR, but with significant complexity savings.

\begin{figure}[t]
	\centering

    \includegraphics[width=0.46\textwidth]{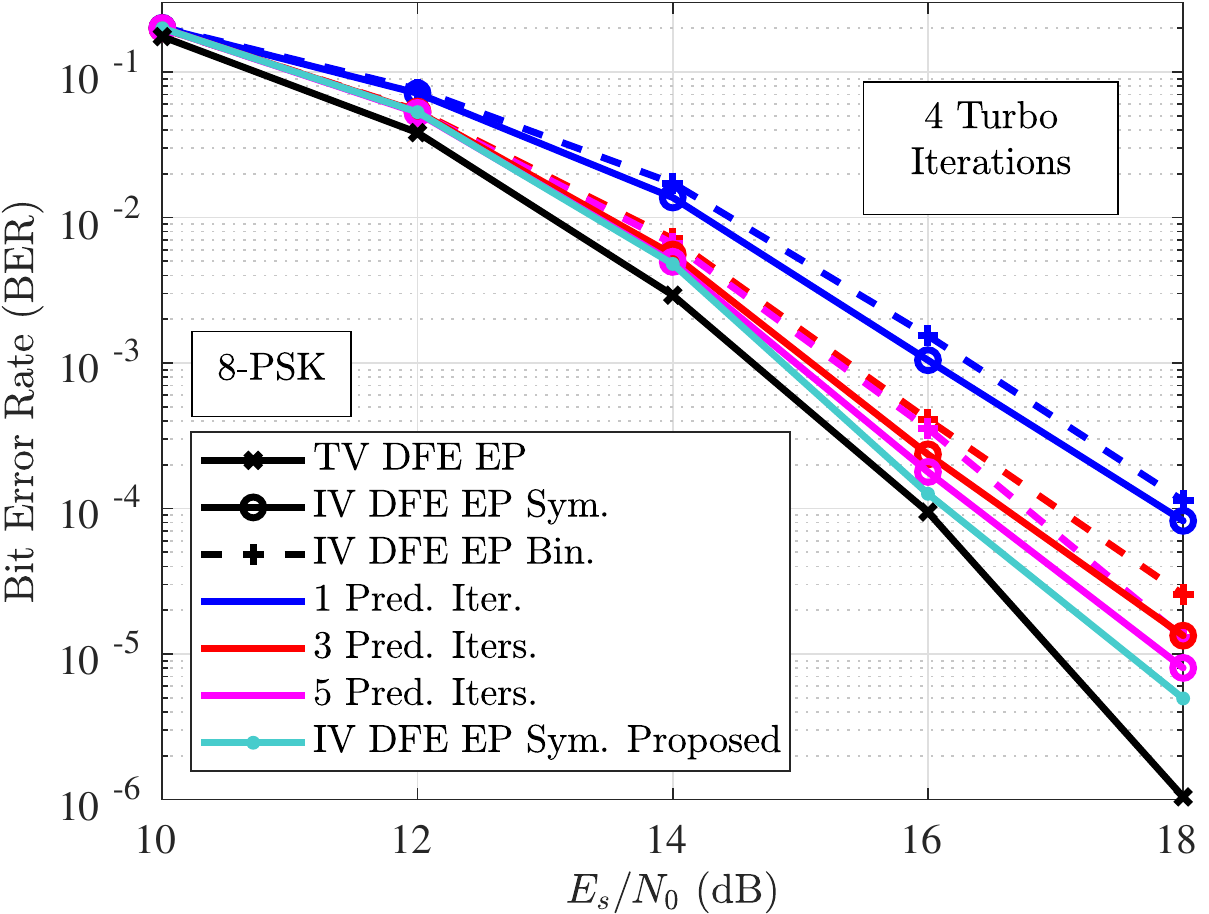}
	\caption{Rate-1/2 coded BER with proposed binary and symbol-wise prediction.
	}
	\label{fig_9_coded_ber_pred_comp}
\end{figure}

\begin{table}[!h]
    \footnotesize
	\renewcommand{\arraystretch}{1.2}
	\caption{Computational Complexity of FIR Receivers}
	\label{table_fir_complexity}
	\centering
	\begin{tabular}{c||c|c}
		\hline
		\bfseries Structure & \bfseries Filter Computation & \bfseries Filtering and Detection \\
		\hline\hline
        %
	    \specialcell{TV LE} & \specialcell{$K(5L^3+56L^2)$} & $K(25L+(11+3q)M)$\\
		\hline
		\specialcell{TV DFE} & \specialcell{$K(5L^3+71L^2)$} & $K(25L+(18+3q)M)$\\
		\hline
		\specialcell{IV LE} & \specialcell{$(6L^3+28L^2)$} & $K(25L+(11+3q)M)$\\
		\hline
		\specialcell{IV DFE} & \specialcell{$(N_\text{pred}+1)(6L^3+34L^2)$} & $K(25L+(18+3q)M)$\\
		\hline
	\end{tabular}
\end{table}

\begin{figure*}[t]
	\centering
    \includegraphics[width=0.95\textwidth]{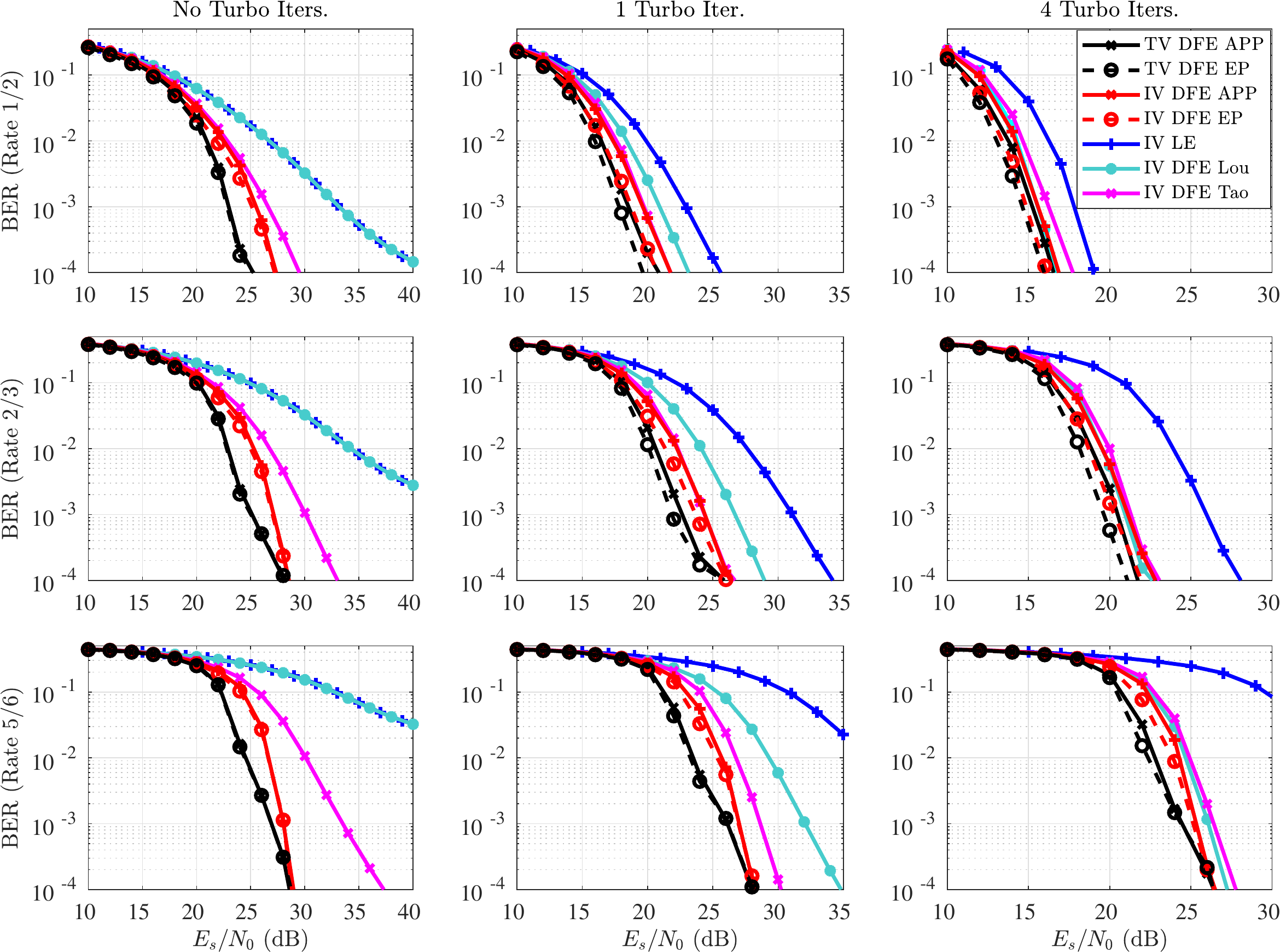}
	\caption{Coded 8-PSK bit-error rate (BER) performance comparison of turbo FIR receivers across turbo-iterations for different code rates.}
	\label{fig_10_coded_ber_all_rates}
\end{figure*}

Approximate computational complexity per turbo-iteration of considered FIR receivers is given in the Table~\ref{table_fir_complexity}.
TV LE and DFE receivers use the reduced-complexity TV matrix inversion algorithms in \cite{sahinCipriano_2018_DFEEP}, and IV receivers exploit Cholesky decomposition for matrix inversion.
The filter computation cost of the proposed IV DFE increases linearly with the number of prediction iterations $N_\text{pred}$.

\subsection{Finite-Length Turbo-Equalization Performance}

In this section, the prediction accuracy is assessed for transmissions encoded with non-recursive non-systematic convolutional code (NRNSCC) of polynomial $[7,5]_8$.

First, the impact of choosing a symbol-wise or a binary prediction scheme is assessed through finite-length BER evaluations. 
The block size is kept at $K=256$, similarly to the uncoded case, and a MAP decoder based on the BCJR algorithm is used as a SISO decoder \cite{bahl_optimal_1974}. 
Fig.~\ref{fig_9_coded_ber_pred_comp} shows the case of the EP-based feedback with 8-PSK, and the use of symbol-wise prediction is shown to accelerate convergence of the IV DFE performance towards TV DFE.

However, despite the improvements brought by the symbol-wise prediction, covariance estimations tend to be too optimistic for high prior information at high SNRs (following 1 or 2 turbo iterations, above 15~dB), and degrade BER performance.
A similar observation was made for the semi-analytic prediction of turbo linear MMSE receivers in \cite{ning_extrinsic_2012}, where a calibration mechanism is applied to correct the predicted prior covariance with a multiplicative penalty factor. 
After some ad hoc optimization, this scheme yields more pessimistic predictions that ends up improving the BER prediction accuracy.

Here, a related mechanism is adapted to the proposed online prediction. 
To avoid over-estimation of the causal covariance, the anti-causal covariance can be exploited to derive a ``lower-bound'' to estimated causal covariances. 
Empirically, turbo detection systems bring most of the improvements at the initial iterations, hence the improvements after a certain number of iterations can no longer be substantial.
Thus, after some trial-and-error tests, we have selected the predicted causal covariance $\bar{v}^c_x$ of the current turbo iteration to be modified with the heuristic $\bar{v}^c_x = \max(\bar{v}^c_x, \beta\bar{v}^a_x)$, with $0 < \beta \le 1$.

\begin{figure}[t]
	\centering

    \includegraphics[width=0.48\textwidth]{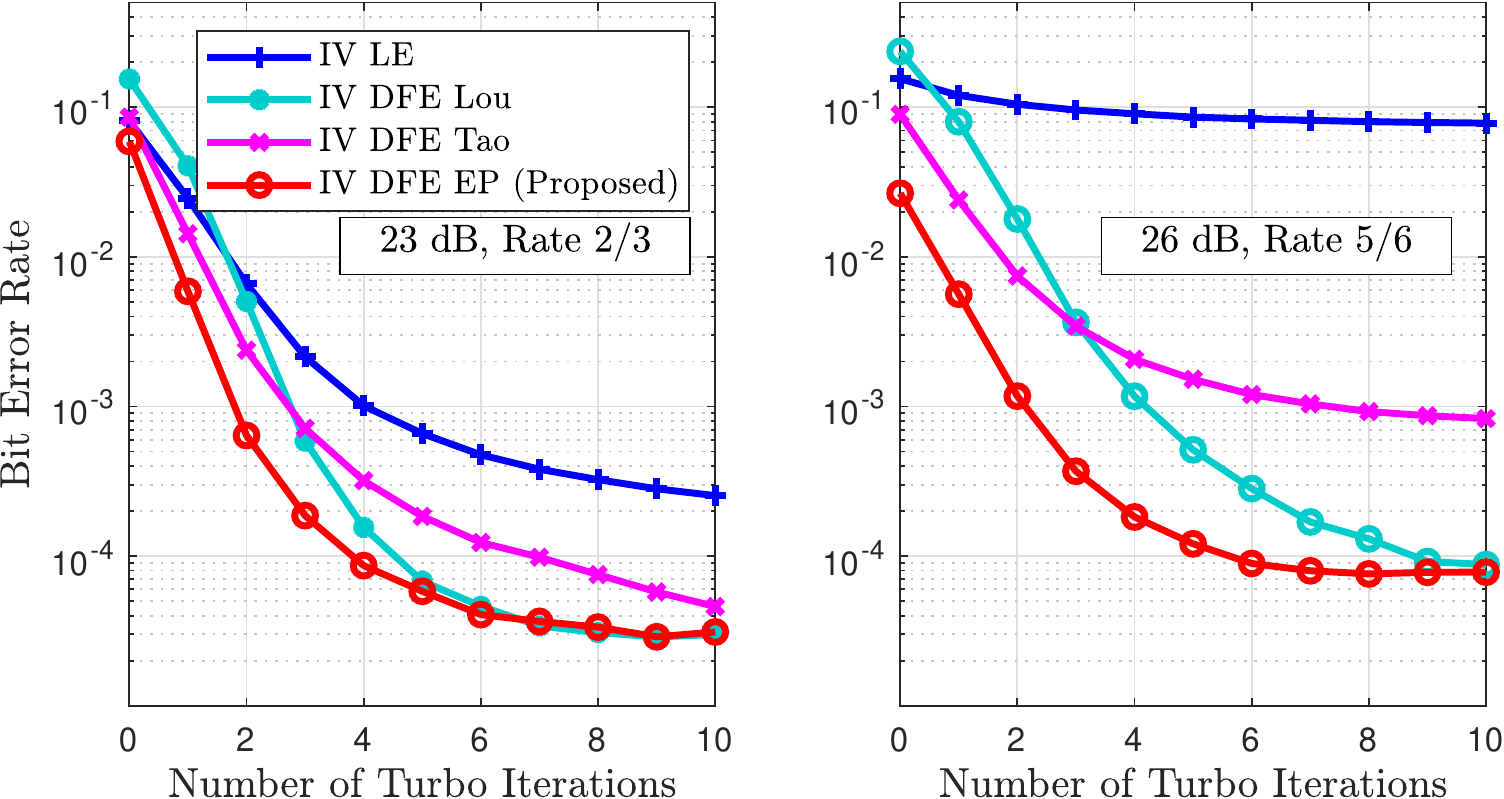}

	\caption{Achievable rates of FIR receivers for 8-PSK in Proakis C channel.
	}
	\label{fig_11_ber_per_turbo_iters}
\end{figure}

The proposed heuristic is integrated with the symbol-wise prediction, with 3 prediction iterations and $\beta=0.2$, and the IV DFE-EP performance is displayed in turquoise in Fig. \ref{fig_9_coded_ber_pred_comp}.

Finally, to compare our proposal to the prior work and to evaluate its behavior in different operating regimes, the previously used rate-1/2 encoding with NRNSCC $[7,5]_8$ is punctured to get rate-2/3 encoding with $[1 1; 0 1]$ puncturing pattern and rate-5/6 encoding with $[1 0 0 0 1; 0 1 1 1 1]$ puncturing pattern.
The BER performance of the proposed IV DFE APP and IV DFE EP receivers are shown in red in Fig.~\ref{fig_10_coded_ber_all_rates}, for 8-PSK transmissions in Proakis C channel, with above mentioned codes of rate 1/2, 2/3 and 5/6, and for 0, 1 and 4 turbo-iterations.
Proposed predictive IV DFE receivers use symbol-wise prediction with 3 iterations, and the heuristic parameter is $\beta=0.2$.
IV DFE APP significantly outperforms other APP-based DFE receivers when there are no turbo iterations, as this is the operating point where the prediction scheme is the most accurate. 
In Figure~\ref{fig_11_ber_per_turbo_iters}, the evolution of BER is plotted as the number of turbo-iterations increases. 
During intermediary iterations of the rate $2/3$ system, previous works of \emph{Tao et al.} \cite{tao_2016_low} and \emph{Lou et al.} \cite{lou_soft_2014} close most of the gap of iteration zero, with the receiver of \cite{lou_soft_2014} slowly converging to the same limit as the proposed receiver.
At high rate systems (rate 5/6) the gap between them and our proposal increases, even for 4 turbo-iterations, and from Figure~\ref{fig_11_ber_per_turbo_iters} it is seen that the receiver of \cite{tao_2016_low} cannot converge to the same asymptotic limits, probably due to the usage of only a few samples for covariance estimation heuristic.
The use of {EP}-feedback instead of {APP} does not bring significant improvement for high-rates, or without turbo-iterations, but at medium and low rates, it allows for an additional asymptotic gain over 0.5~dB.
However, the predictability of the {EP} feedback over a wider set of configurations makes it a more attractive solution.

 \section{Conclusion}

This paper carries out an original approach to the design of turbo DFE receivers with static filters, through the use of online prediction, based on semi-analytic performance prediction techniques as used in physical layer abstraction methods.
Due to the lack of a closed-form solution for such receivers, various heuristics are used throughout the literature. However, discussion on the optimality of such approaches was lacking and it is one of the contribution of this paper.

Here, semi-analytical performance prediction of exact time-varying turbo DFE with dynamic filters is exploited to derive static DFE filters. 
This approach has been carried out for DFE with APP-based or EP-based soft feedback and their detection performance has been evaluated in various configurations.
This framework could also be applied to self-iterated FIR DFE \cite{sahinCipriano_2018_DFEEP} for further improved performance, by updating anti-causal variances with causal EP variance of the last self-iteration.

Our analysis shows that significant complexity savings can be achieved with respect to TV DFE, while offering reasonably close performance. 
Moreover, our method is compatible with any constellation, and spectrally efficient on a large interval of coding rates and with or without turbo-iterations.

	\ifCLASSOPTIONcaptionsoff
	\newpage
	\fi

	
	
	
	\bibliographystyle{IEEEtran}
	\bibliography{IEEEabrv,bibDFE}

\ifdouble
\begin{IEEEbiography}[{\includegraphics[width=1in,height=1.25in,clip,keepaspectratio]{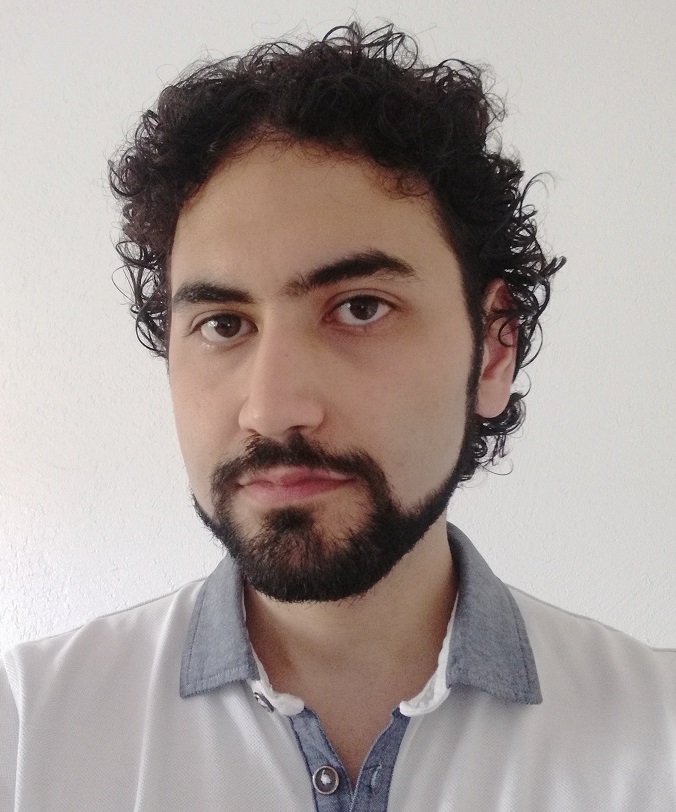}}]{Serdar \c{S}ahin}
	was born in Ankara, Turkey, in 1992.
	He received the M.Sc.Eng. degree in control sys-tems and electronics engineering from INSA de Toulouse, University of Toulouse, France, in 2015.
	He is currently pursuing the PhD degree in digital communications with IRIT-ENSEEIHT, Toulouse, and also with Thales Communications and Security, Gennevilliers. 
	His main research interests include iterative receiver design, practical cooperative transmission schemes and PHY layer abstraction.
\end{IEEEbiography}

\begin{IEEEbiography}[{\includegraphics[width=1in,height=1.25in,clip,keepaspectratio]{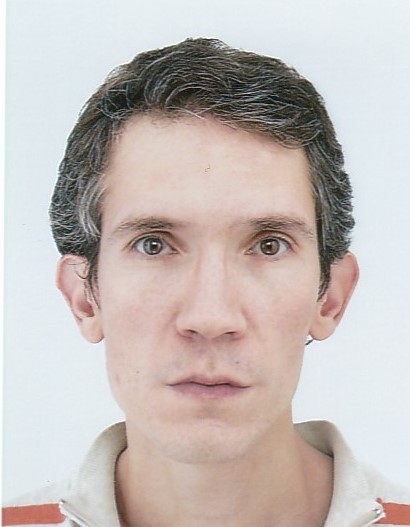}}]{Antonio Maria Cipriano} was born in Padova, Italy, in 1976. He received the Laurea degree in telecommunications engineering from the University of Padova, Italy, in 2000, and the Ph.D. degree in digital communications jointly from the University of Padova and the Ecole Nationale Sup\'{e}rieure des T\'{e}l\'{e}communications (ENST) Paris, France, in 2005. 
In 2001, he was a Young Engineer at Eutelsat, France, for eight months. 
From 2005 to 2007, he held two post-doctoral positions at ENST, Paris, and Orange Labs. 
In 2007 he joined Thales Communication and Security as a digital communication engineer and was involved in several national and international research projects on 4G and 5G communication systems. 
His main research interests lie in the broad area of digital communication systems. 
He is currently involved in research about PHY layer abstractions, relaying for ad hoc mobile networks and advanced receiver design.
\end{IEEEbiography}

\begin{IEEEbiography}[{\includegraphics[width=1in,height=1.25in,clip,keepaspectratio]{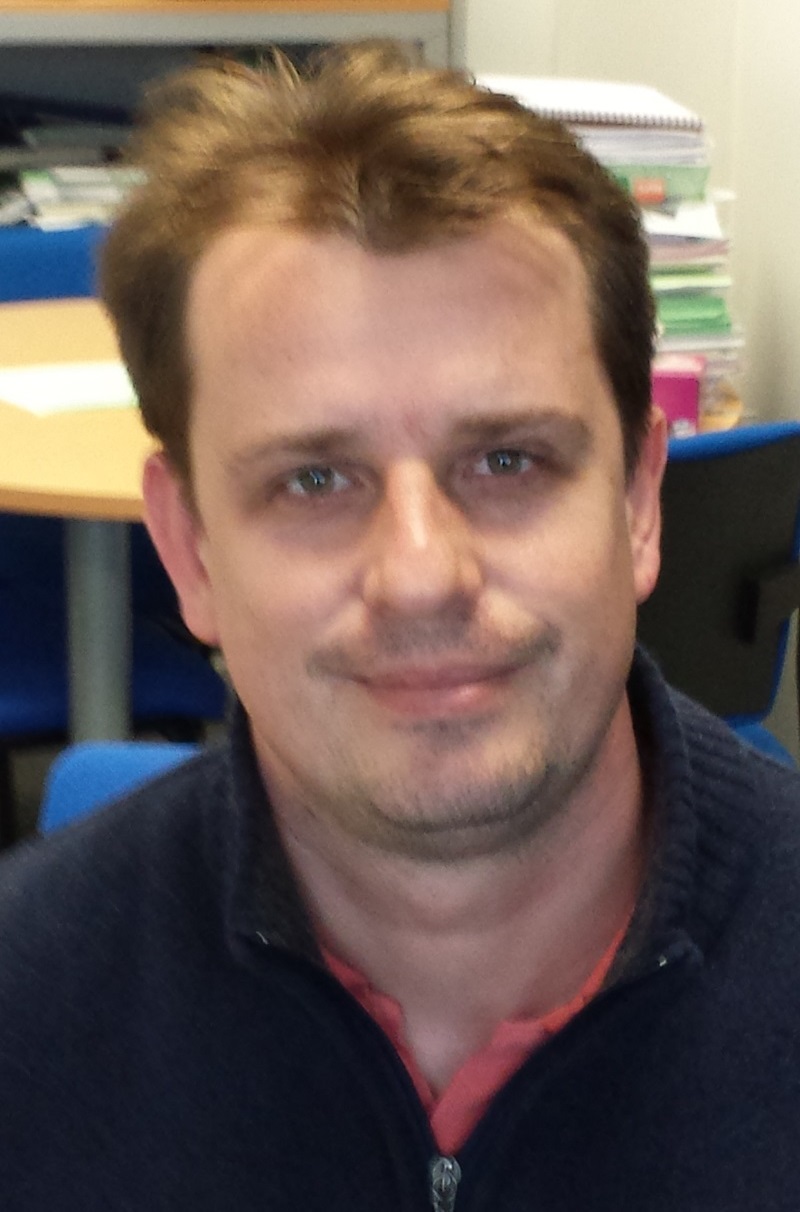}}]{Charly Poulliat}
	received the M.Sc.Eng. degree in electrical engineering from the Ecole Nationale Sup\'{e}rieure de l'Electronique et de ses Applications (ENSEA), Cergy-Pontoise, France, in 2001 and the M.S. degree in Image and Signal Processing from the University of Cergy-Pontoise, France, in 2001, and his PhD degree in electrical and computer engineering from the University of
	Cergy-Pontoise, France, in 2004, and the Habilitation degree from the University of Cergy-Pontoise in 2010.
	From 2004 to 2005, he was a Post-Doctoral Researcher at UH Coding Group, University of Hawaii at Manoa, HI, USA, supervised by Prof. M. Fossorier.
	In 2005, he joined the Signal and Telecommunications department of the engineering school ENSEA as an Assistant Professor. 
	Since 2011, he has been a Professor with the
	National Polytechnic Institute of Toulouse (INP-ENSEEIHT), University of Toulouse. He is also with the Signal and Communications Group, CNRS IRIT Laboratory.
	His research interests include signal processing for digital communications, waveform design, channel coding,
	iterative system design and optimization.
\end{IEEEbiography}

\begin{IEEEbiography}[{\includegraphics[width=1in,height=1.25in,clip,keepaspectratio]{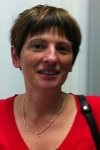}}]{Marie-Laure Boucheret}
	received the M.Sc.Eng. degree in electrical engineering from ENST Bretagne, Brest, France in 1985, the Ph.D. degree in Digital Communications from TELECOM ParisTech, in 1997, and the Habilitation \`{a} diriger les recherches degree from INPT University of Toulouse, in 1999. 
	From 1985 to 1986, she was a Research Engineer with the French Philips Research Laboratory (LEP). From 1986 to 1991, she was an Engineer with Thales Alenia Space, first as a Project Engineer (TELECOM II program) then as a Study Engineer at the Transmission Laboratory. 
	From 1991 to 2005, she was with TELECOM ParisTech first as an Associated Professor then as a Professor. 
	Since 2005, she has been a Professor with the National Polytechnic Institute of Toulouse (INP-ENSEEIHT),  
	University of Toulouse.  She is also with the Signal and Communication Group, IRIT Laboratory. Her fields of interest are signal processing for communication and satellite communications.
\end{IEEEbiography}
\else
\fi

\end{document}